\documentclass{iaus}

\usepackage{graphicx}

\title[Nucleosynthesis results from INTEGRAL] %% give here short title %%
{Recent nucleosynthesis results from INTEGRAL}

\author[G. Weidenspointner]   %% give here short author list %%
{G. Weidenspointner}

\affiliation{Centre d'Etude Spatiale des Rayonnements, 9, avenue
du Colonel-Roche, BP 4346, 31028 Toulouse Cedex 4, France \break
email: Georg.Weidenspointner@cesr.fr}

\pubyear{2005}
\volume{230}  %% insert here IAU Symposium No.
\pagerange{119--126}
\date{?? and in revised form ??}

\jname{``Populations of High Energy Sources in Galaxies''}

\editors{Evert J.A. Meurs \& G. Fabbiano, eds.}

\begin{document}

\maketitle

\begin{abstract}

Since its launch in October 2002, ESA's INTEGRAL observatory has
enabled significant advances to be made in the study of Galactic
nucleosynthesis. In particular, the imaging Ge spectrometer SPI
combines for the first time the diagnostic powers of high
resolution gamma-ray line spectroscopy and moderate spatial
resolution. This review summarizes the major nucleosynthesis results
obtained with INTEGRAL so far. Positron annihilation
in our Galaxy is being studied in unprecented detail. SPI observations
%allowed us to obtain 
yield the first
%all-
sky maps in both the 511~keV annihilation line and the positronium
continuum emission, and 
%to perform 
the most accurate spectrum at 511~keV to
date, thereby imposing new constraints on the source(s) of Galactic
positrons which still remain(s) unidentified. For the first time, the
imprint of Galactic rotation on the centroid and shape of the 1809~keV
gamma-ray line due to the decay of $^{26}$Al has been seen, confirming
the Galactic origin of this emission.
%In addition, detailed spectroscopy
%of 1809~keV line emission from the Cygnus region provides new insights
%into the dynamics of the interstellar medium in this important star
%forming region. 
SPI also provided the most accurate determination of
the gamma-ray line flux due to the decay of $^{60}$Fe. The combined results
for $^{26}$Al and $^{60}$Fe have important implications for
nucleosynthesis in massive stars, in particular Wolf-Rayet
stars. 
%Both IBIS and SPI have detected gamma-ray line emission
%associated with $^{44}$Ti from the supernova remnant Cas~A, with
%interesting implications on the physics of core-collapse
%supernovae. Both instrumenst 
Both IBIS and SPI are searching the Galactic plane for young supernova
remnants emitting the gamma-ray lines associated with radioactive
$^{44}$Ti. None have been found so far,
%which may call into question the correctness of the commonly
%accepted Galactic supernova rate of 2--3 per century.
which raises important questions concerning the production of
$^{44}$Ti in supernovae, the Galactic supernova rate, and the Galaxy's
chemical evolution.



\end{abstract}

\firstsection % if your document starts with a section,

%=======================================================================

\section{Introduction}\label{intro}

%Ever since the advent of gamma-ray astronomy, nucleosynthesis theory
%and gamma-ray spectroscopy have greatly benefited from each other.

Gamma-ray line astronomy has opened a new and unique window for
studying nucleosynthesis in our Galaxy. The singular advantage of
gamma-ray spectroscopy over other observations is that it offers the
opportunity to detect directly and identify uniquely individual
isotopes at their birthplaces.
%The characteristic gamma-ray lines emitted in radioactive
%decays allow individual isotopes to be uniquely identified.
%, since the
%gamma-ray lines reflect transitions in the atomic nucleus, unlike
%transitions in the atoms' electron shell observed at lower
%energies. 
The rate at which radioactive decays proceed is in general
unaffected by the physical conditions in their environment, such as
temperature or density.
%\footnote{A potential exception could be
%electron-capture decay in case of temperatures high enough to strip
%practically all electrons off the nucleus.}. 
The interstellar medium is
not dense enough to attenuate gamma rays, so that radioactive decays
can be observed throughout our Galaxy. Recent reviews on implications
of gamma-ray observations for nucleosynthesis in our Galaxy can be
found in \cite{D-P-vB05} and \cite{Prantzos05}.

%In this invited contribution, I will present a selection of recent
%nucleosynthesis results obtained with ESA's INTEGRAL
%observatory and discuss their astrophysical significance.
%Specifically, I will address observations concerning the radioisotopes
%$^{44}$Ti, $^{26}$Al, and $^{60}$Fe, and the annihilation of positrons
%in our Galaxy. 
%To end, I will summarize the status of gamma-ray line
%astronomy early into the INTEGRAL mission before providing some of the
%propects that the continued operation of this powerful observatory may
%hold.

%=======================================================================

\section{Results}\label{results}

The nucleosynthesis results presented in the following have all been
obtained from observations with the two main instruments on board the
INTEGRAL observatory: the spectrometer SPI and the imager IBIS (for
details regarding the instruments, see \cite{Kretschmar05} and
references therein). These two instruments are complementary in their
%performance 
characteristics, 
%thereby 
providing an unprecedented view of the Universe at hard X-ray and soft
gamma-ray energies. The imaging Ge spectrometer SPI offers high
spectral resolution of about 2.1~keV FWHM at 511~keV combined for the
first time with moderate spatial resolution (FWHM about
$3^\circ$). The imager IBIS offers excellent spatial resolution of
about $12^\prime$ FWHM at moderate spectral resolution (FWHM about
38~keV at 511~keV).

%-----------------------------------------------------------------------

\subsection{$^{44}$Ti}\label{44Ti}

%-----------------------------------------------------------------------

\subsubsection{$^{44}$Ti production sites and
processes}\label{44Ti_prod}

%The radioisotope $^{44}$Ti is mainly produced by massive stars in
%their final explosion as core-collapse supernovae (ccSNe) 

The radioisotope $^{44}$Ti is primarily produced in the so-called
$\alpha$-rich freeze-out of material initially in nuclear statistical
equilibrium. 
%As the material expands more and more reactions
%fall out of equilibrium. In particular, remaining free $\alpha$
%particles can no longer combine to $^{12}$C nuclei to eventually merge
%back into Fe group nuclei since the triple-$\alpha$ reaction is very
%sensitive to density; they are therefore available for capture on
%remaining heavier nuclei (\cite{The98}). 
The main site for
$\alpha$-rich freeze-out to occur is thought to be the innermost
layers of core-collapse supernovae (ccSNe), although sub-Chandrasekhar
mass white dwarf Type~Ia SNe have also been proposed (\cite[Woosley \& Weaver 1994]{Woosley_Weaver94}).

\begin{figure}
%\centerline{\includegraphics[width=8cm,bbllx=68pt,bblly=372pt,bburx=540pt,bbury=694pt,clip=]{renaud.yield_vs_rate_gauss2.ps}}
%\centerline{
%\includegraphics[width=5cm,angle=270.]{Prantzos_f2.eps}
%\includegraphics[width=5cm,angle=270.]{Prantzos_f7.eps}
%}
\centerline{
\includegraphics[width=7cm]{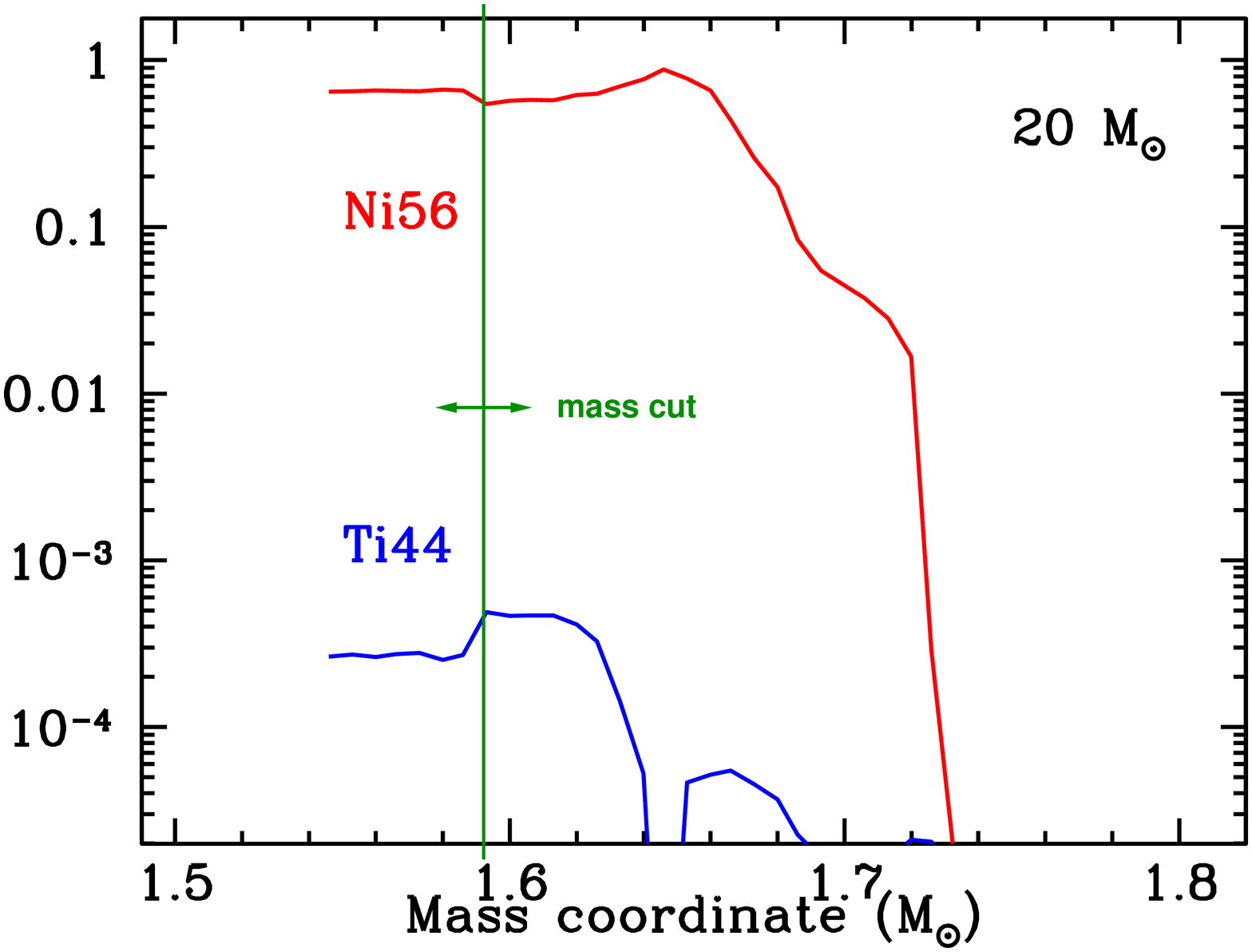}
\includegraphics[width=7cm]{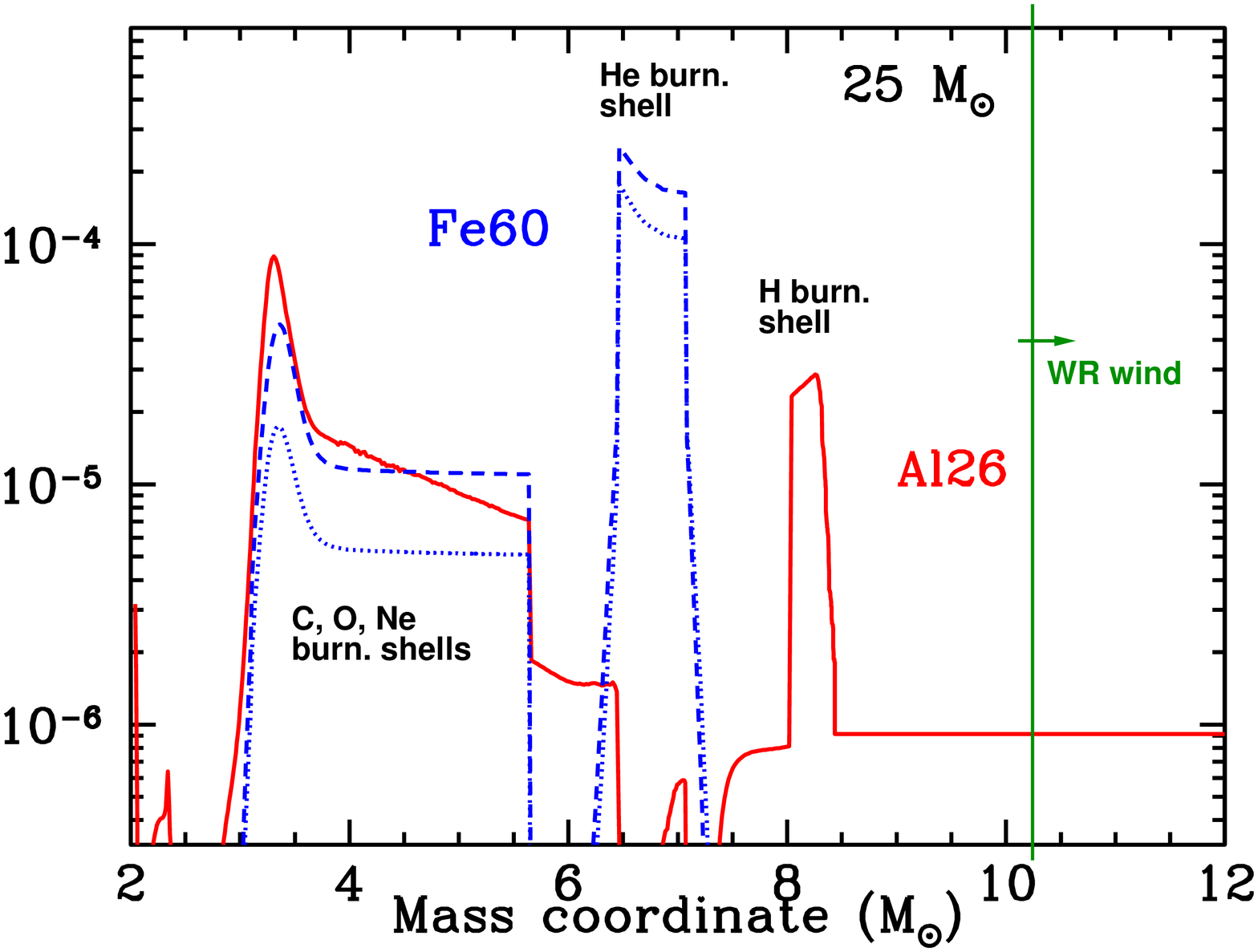}
}
\caption{Left panel: radial abundance profiles (mass fractions) of
$^{56}$Ni and $^{44}$Ti inside a 20~M$_\odot$ star after the passage
of the shock front. Right panel: radial abundance profiles (mass
fractions) of $^{26}$Al and $^{60}$Fe inside a 25~M$_\odot$ star after
the passage of the shock front. Both figures were adapted from
\cite{Prantzos05}.}
\label{prantzos_44Tifig}
\end{figure}

The $^{44}$Ti yield of ccSNe is notoriously difficult to calculate
because it depends sensitively on the so-called mass cut, the
explosion energy, and the (a)symmetry of the explosion. The mass cut,
which has not yet been successfully calculated and is illustrated in
the left panel of Fig.~\ref{prantzos_44Tifig}, is the notional surface
separating material that is ejected from material that will fall back
onto the compact remnant (neutron star or black hole) of the
explosion. $^{44}$Ti is believed to be produced in the deepest layers
of the exploding star that may be ejected, depending on the precise
location of the mass cut. The amount of synthesized $^{44}$Ti also
depends sensitively on the explosion energy and (a)symmetry.
%, as both
%strongly affect the rate at which the temperature and density of the
%ejecta are dropping. 
Theoretical calculations indicate that both
increased explosion energy and increased asymmetry 
% tend to
result in an increased $^{44}$Ti yield.

Observationally, the presence of the radioisotope $^{44}$Ti is
revealed to the gamma-ray astronomer through the emission of three
gamma-ray lines. The decay $^{44}$Ti~$\rightarrow$~$^{44}$Sc
($\tau_{^{44}Ti} \sim 90$~y) gives rise to gamma rays at 67.9~keV and
78.4~keV; the subsequent decay $^{44}$Sc~$\rightarrow$~$^{44}$Ca
($\tau_{^{44}Sc} \sim 5.7$~h) gives
rise to a line at 1157.0~keV.

The astrophysical interest in $^{44}$Ti is two-fold. Clearly, the
amount and the velocity of $^{44}$Ti is a very powerful probe of the
explosion mechanism and dynamics of ccSNe, which are still poorly
understood. In addition, the $^{44}$Ti gamma-ray line emission
% due to the decay of
%$^{44}$Ti 
is an ideal indicator of young SN remnants (SNRs). The lifetime 
%of $^{44}$Ti 
is about 90~y, which roughly coincides with the expected recurrence
time interval for ccSNe in our Galaxy.
% of about 30--100~y. 
It is
therefore expected that
%, provided the availability of 
with a sufficiently sensitive instrument a few young SNRs should be
visible in our Galaxy at the current epoch.

%-----------------------------------------------------------------------

%\subsubsection{Cas~A}\label{44Ti_CasA}
%
%{\bf ??? skip ???}

%-----------------------------------------------------------------------

\subsubsection{Search for young SNRs in the inner
Galaxy}\label{44Ti_SNRsearch}

%As I pointed out above, the gamma-ray line emission due to the decay
%of $^{44}$Ti is an ideal signature of young SNRs. 
The most sensitive search to date for young SNRs at gamma-ray energies
was performed by \cite{Renaud04} who used the first year of INTEGRAL
observations to search for 68~keV and 78~keV line emission in the
inner Galaxy with the imager IBIS.  This search addresses a
long-standing puzzle linking the Galactic SN rate and Galactic chemical
evolution: given current estimates of the present-day rates of
thermonuclear and ccSNe and their yields, these events can only
account for about $1/3$ of the solar $^{44}$Ca abundance based on
chemical evolution models and assuming that all $^{44}$Ca is formed as
$^{44}$Ti (\cite[Leising \& Share 1994]{Leising_Share94}). At the
same time, given these SN properties, combined with models for their
Galactic distribution, past missions should have detected a few young
SNRs even with their lower sensitivities -- and detections were
certainly expected for the unprecedented limiting point source
sensitivity
%($3\sigma$) of about $1.8\times10^{-5}$~ph~cm$^{-2}$~s$^{-1}$ 
achieved with IBIS.
% so far.
However, as was the case in less sensitive previous searches, none
have been found.
% -- deepending the puzzle of the origin of Galactic
%$^{44}$Ca.

\begin{figure}[t]
\begin{minipage}[b]{7.cm}
\includegraphics[width=7cm,bbllx=68pt,bblly=372pt,bburx=540pt,bbury=694pt,clip=]{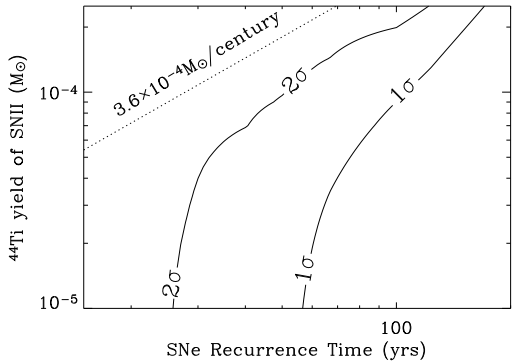}
\end{minipage}
\hfill
%\begin{minipage}[t]{5.0cm}
%\vspace*{-5.6cm}
\begin{minipage}[t]{6.0cm}
\vspace*{-4.9cm}
\caption{Contours denoting the probability that IBIS/ISGRI would have
detected at least one $^{44}$Ti point source as a function of the
$^{44}$Ti yield of Type~II SNe and the Type~II SN recurrence time
(figure adapted from \cite[Renaud \etal\ 2004]{Renaud04}). The dotted
line represents the present-day $^{44}$Ti production rate that is
required to explain the solar system abundance of $^{44}$Ca, assuming
that is solely produced as $^{44}$Ti in Type~II SNe.}
\label{renaud_fig}
\end{minipage}
\end{figure}

To assess the implications of the non-detection of young SNRs,
\cite{Renaud04} estimated the probability that at least one $^{44}$Ti
point source is detectable
% in the inner Galaxy by performing a Monte Carlo study. 
by generating Monte Carlo distributions using current estimates of the
rates of thermonuclear and ccSNe, of their yields, and of their
Galactic distribution; SN explosions were simulated as a function of 
%in space, time, and $^{44}$Ti yields; the 
the recurrence time and the $^{44}$Ti yield of
Type~II SNe. A typical result is depicted in
Fig.~\ref{renaud_fig}.

%\begin{figure}
%\centerline{\includegraphics[width=8cm,bbllx=68pt,bblly=372pt,bburx=540pt,bbury=694pt,clip=]{renaud.yield_vs_rate_gauss2.ps}}
%\caption{Contours denoting the probability that IBIS/ISGRI would have
%detected at least one $^{44}$Ti point source as a function of the
%$^{44}$Ti yield of Type~II SNe and the Type~II SN recurrence time
%(figure adapted from \cite{Renaud04}). The dotted line represents the
%present-day $^{44}$Ti production rate that is required to explain the
%solar system abundance of $^{44}$Ca, assuming that is solely produced
%as $^{44}$Ti in Type~II SNe.}\label{renaud_fig}
%\end{figure}

The Monte Carlo study rules out two obvious solutions to the $^{44}$Ca
puzzle. The estimated Galactic $^{44}$Ti production rate of ccSNe
could in principle be increased by postulating that the Galactic SN
rate has been underestimated, or by postulating that the $^{44}$Ti
yields have been underestimated. Neither solution is viable, because
in order to account for the solar $^{44}$Ca abundance by SNe a
present-day production rate of $3.6 \times
10^{-4}$~M$_\odot$~century$^{-1}$ is required -- which is excluded at
a greater than 95\% confidence level for any acceptable combination of
SN rates and $^{44}$Ti yields. A third solution, compatible with
existing data, is the existence of another, rare but high yield,
Galactic $^{44}$Ti source such as sub-Chandrasekhar mass white dwarf
Type~Ia SNe. The latest search for young SNRs with IBIS therefore
highlights important open questions concerning the chemical evolution
of our Galaxy, the production of $^{44}$Ti in SNe, and the Galactic SN
rate (\cite[Renaud \etal\ 2004]{Renaud04}).

%-----------------------------------------------------------------------

\subsection{$^{26}$Al and $^{60}$Fe}\label{26Al-60Fe}

%-----------------------------------------------------------------------

\subsubsection{$^{26}$Al and $^{60}$Fe production sites and
processes}\label{26Al-60Fe_prod} 

$^{26}$Al and $^{60}$Fe are produced hydrostatically as well as
explosively in massive stars. The production of $^{26}$Al requires
free protons, hence it is mainly produced in the H, Ne, and C burning
layers; the production of $^{60}$Fe requires free neutrons, hence it
is mainly produced in C, Ne, and He burning layers, as indicated in
the right panel of Fig.~\ref{prantzos_44Tifig}. Products of
hydrostatic nucleosynthesis
% (such as $^{26}$Al in the H burning shell)
can escape the massive star through
stellar winds, in particular during the Wolf-Rayet (WR) phase of the
most massive stars. The $^{26}$Al, being produced in H burning layers
near the surface, can escape most easily.
%At first, only $^{26}$Al is expelled. In rare
%cases, eventually $^{60}$Fe can be expelled as well. 
When the star explodes in a ccSN, about equal amounts of $^{26}$Al and
$^{60}$Fe are ejected. The yields of $^{60}$Fe and in particular of
$^{26}$Al are hard to predict because of the important role of stellar
winds, which in turn depend e.g.\ on the metallicity of the star or
its rotation.

Both radioisotopes can be observed by gamma-ray line spectroscopy. The
decay of $^{26}$Al ($\tau_{^{26}Al} \sim 1.1\times10^6$~y) gives rise
to a single gamma-ray line at 1808.6~keV. The decay
$^{60}$Fe~$\rightarrow$~$^{60}$Co ($\tau_{^{60}Fe} \sim
2.2\times10^6$~y) is revealed by a line at 58.6~keV, the subsequent
decay $^{60}$Co~$\rightarrow$~$^{60}$Ni ($\tau_{^{60}Co} \sim 7.6$~y)
is revealed by two lines at 1173.2~keV and 1332.5~keV.

Astrophysically, the subtle differences in the production zones of the
radioisotopes $^{26}$Al and $^{60}$Fe render their combined study an
important probe of massive star nucleosynthesis. Furthermore, since the
lifetimes of these two radioisotopes are 
%much longer than the characteristic time interval of about 100~y
%between individual Galactic ccSNe, but 
much shorter than Galactic evolution timescales but are large enough to
accumulate in the ISM, $^{26}$Al and
$^{60}$Fe are ideally suited to study the sites and the distribution
of recent Galactic star formation.

%-----------------------------------------------------------------------

%\subsubsection{$^{26}$Al in Cygnus~X star forming
%region}\label{26Al_CygX}
%
%{\bf ??? skip ???}

%-----------------------------------------------------------------------

\subsubsection{$^{26}$Al in the inner Galaxy}\label{26Al_innerGalaxy}

%$^{26}$Al is the first radioactive isotope detected by gamma-ray
%spectroscopy (\cite{Mahoney84}). 
The sky distribution of Galactic $^{26}$Al was first mapped using the
COMPTEL instrument; it was found to be very similar to that of massive
stars in our Galaxy, establishing the massive star origin of $^{26}$Al
(\cite[Kn\"odlseder \etal\ 1999]{Knoedlseder99}). However, the
relative importance of WR star winds and of ccSNe for seeding the ISM
remained unclear. Early high-resolution spectroscopy with the balloon
borne GRIS instrument by
\cite{Naya96} yielded an unexpectedly large line width of about
5.4~keV FWHM and complicated rather than clarified the origin of
$^{26}$Al as such substantial line broadening could only be explained
by rather unusual circumstances (\cite[Sturner \& Naya
1999]{Sturner_Naya99}). Spatially resolved spectroscopy, as afforded
by the imaging spectrometer SPI, is expected to resolve the issue, as
it will make it possible to probe the imprint of the dynamics of the
ISM local to the $^{26}$Al sources and that of Galactic differential
rotation on shape and centroid of the 1808.6~keV line (modelled e.g.\
by \cite[Kretschmer \etal\ 2003]{Kretschmer03}).

\begin{figure}
\centerline{
\includegraphics[width=7cm]{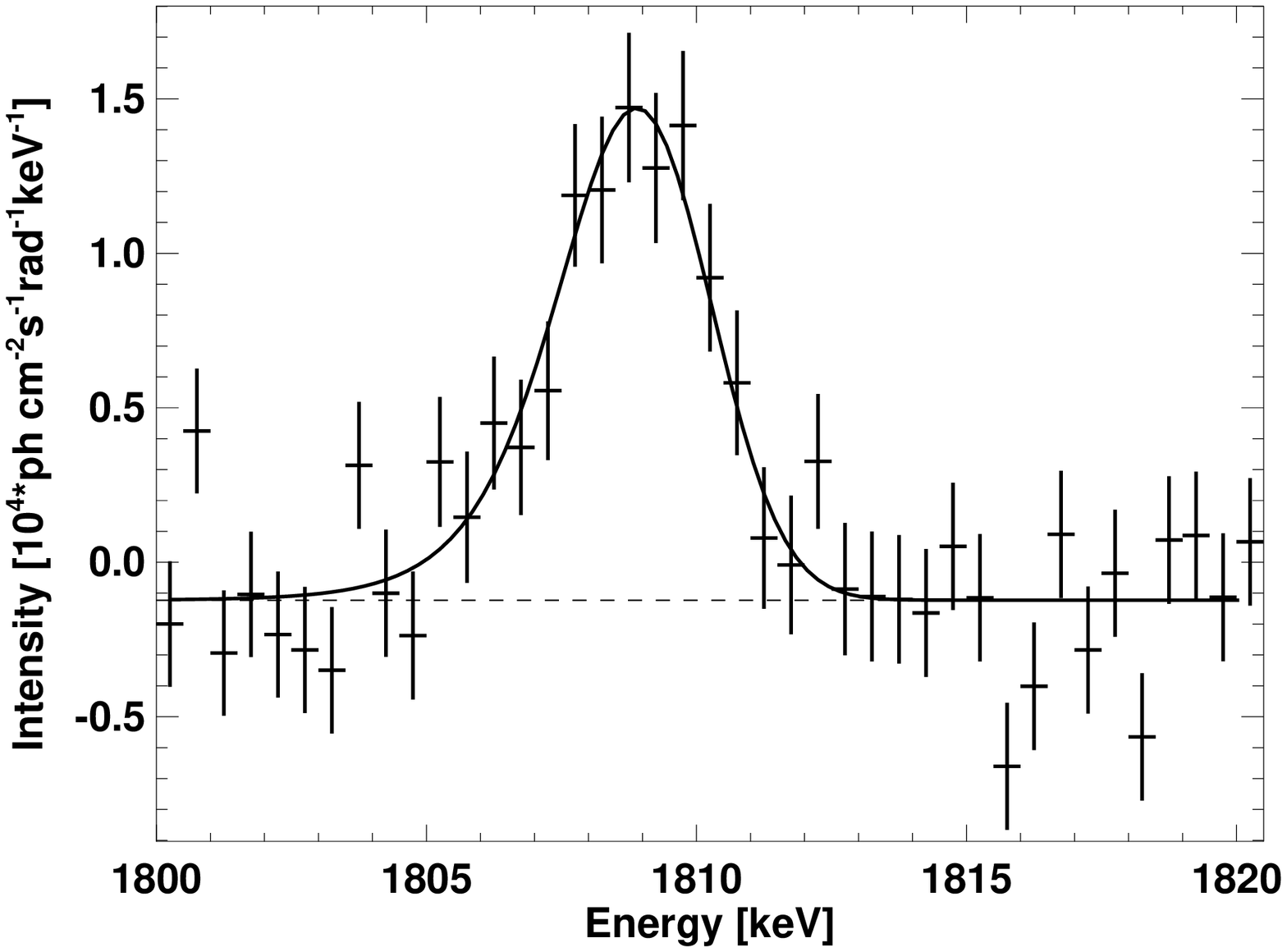}
\includegraphics[width=7cm]{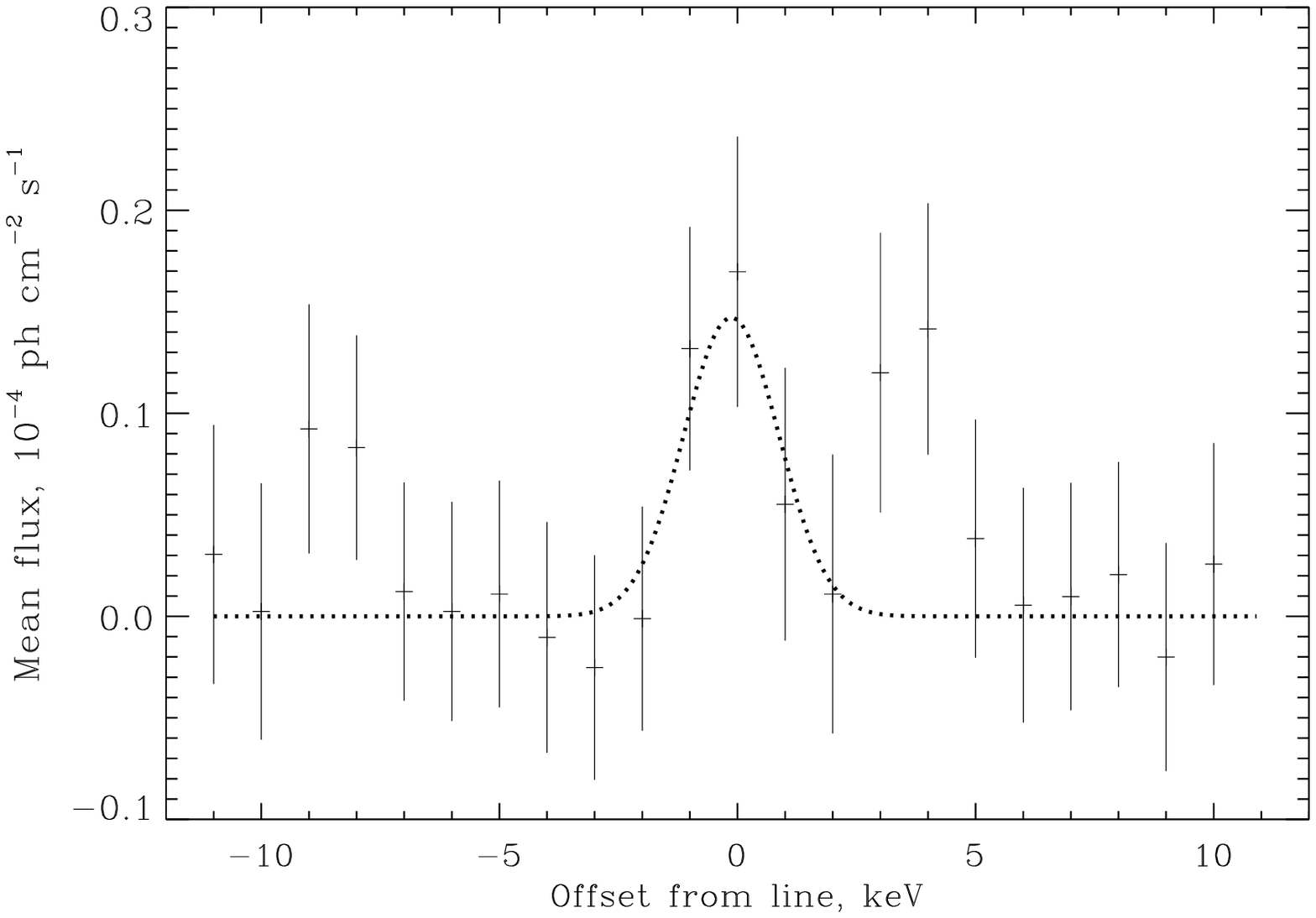}
}
%\caption{Left panel: the spectrum of the $^{26}$Al gamma-ray line
%emission from the inner Galaxy as obtained by \cite{Diehl05b}. Right
%panel: the overlayed and summed spectrum of the observed lines at
%1173.2~keV and 1332.5~keV (\cite[Harris \etal\ 2005]{Harris05}).}
\caption{Left panel: the spectrum of the $^{26}$Al gamma-ray line
emission from the inner Galaxy (data points) and a Gaussian line fit
(solid line) as obtained by \cite{Diehl05b}.
Right panel: the overlayed and summed spectrum of the observed
$^{60}$Fe lines at 1173.2~keV and 1332.5~keV (data points) and
a Gaussian line fit (dashed line) according to \cite{Harris05}.}
\label{26Al_60Fe_spectra}
\end{figure}

Using the first 1.5~years of observations with SPI,
\cite[Diehl \etal\ (2005a,b)]{Diehl05a,Diehl05b} obtained a spectrum
of the 1808.6~keV line from the inner Galaxy of unprecedented quality
(see left panel of Fig.~\ref{26Al_60Fe_spectra}). The line centroid is
unshifted; the line width is intrinsically narrow with a FHWM of about
1.2~keV (and a 2$\sigma$ upper limit of 2.8~keV), ruling out the
earlier GRIS result. These line parameters are consistent with
expectations from the
\cite{Kretschmer03} model, which for the inner Galaxy predicts no line
shift and a line width of about 1~keV due to Galactic differential
rotation (leaving room for a small amount of additional broadening due
to modest ISM turbulence around the $^{26}$Al sources). The $^{26}$Al
line is detected with such high significance by SPI that spatially
resolved spectroscopy in the inner Galaxy can be performed for the
first time, with important implications for the amount and origin of
$^{26}$Al, and the Galactic star formation and SN rates as discussed
by \cite{Diehl05a}.

%-----------------------------------------------------------------------

\subsubsection{$^{60}$Fe and $^{26}$Al from massive
stars}\label{26Al-60Fe_massivestars}

%As pointed out earlier, $^{26}$Al and $^{60}$Fe are produced under
%similar, but not quite the same, conditions in massive
%stars. Understanding the nucleosynthesis of one of the isotopes is
%greatly helped by any information that can be obtained on the
%other. 
One of the great successes for SPI 
%therefore 
was the first
significant measurement of $^{60}$Fe emission from our Galaxy by
\cite{Harris05} using the observations of the first year. The summed
and overlayed spectrum of the two observed lines 
%at 1173.2~keV and 1332.5~keV 
is depicted in the right panel of
Fig.~\ref{26Al_60Fe_spectra}. The observed flux per line is $(3.7 \pm
1.1) \times 10^{-5}$~ph~cm$^{-2}$~s$^{-1}$, which translates into a
line flux ratio of $^{60}$Fe to $^{26}$Al of $0.11
\pm 0.03$ per $^{60}$Fe line.

The detection of $^{60}$Fe has important implications for the
nucleosynthesis in massive stars, as was first pointed out by
\cite[Prantzos (2004, 2005)]{Prantzos04, Prantzos05}. He integrated
the predicted yields for Type~II SNe over an initial mass function and
found that if he only considered the lower mass range, where the WR
phase is absent or weak, the $^{60}$Fe to $^{26}$Al line flux ratio is
significantly overpredicted. This means that there must be an
additional $^{26}$Al source; Type~II SNe do not produce enough
$^{26}$Al. This additional source could be WR stars, because when the
integration is extended up to the highest masses, taking into account
the $^{26}$Al expelled in the WR winds, the observed $^{60}$Fe to
$^{26}$Al line flux ratio can be reproduced.  However, one should keep
in mind that WR star yields are still uncertain, as they depend
sensitively on e.g.\ metallicity and rotation.

%-----------------------------------------------------------------------

\subsection{Galactic positrons}\label{positrons}

\subsubsection{Production processes and positron
sources}\label{pos_prod}

%Cosmic positron annihilation radiation was first detected in
%balloon observations from the Galactic centre (GC) direction in the 1970s
%and has been the focus of intense scrutiny by a large number of
%balloon and satellite borne experiments ever since.
%%%\citep[see reviews by, e.g.,][]{lingenfelter_ramaty89, harris97}.
%%%\citep[see e.g.\ the reviews by][]{Tueller92,Harris97}.
%%%(see e.g.\ the reviews by \cite{Tueller92} and \cite{Harris97}).
%%(see e.g.\ the review by \cite[Harris 1997]{Harris97}).
%%%
%Despite much observational and theoretical effort, the origin
%of the positrons is still poorly understood.

The annihilation of positrons with electrons gives rise to two
characteristic emissions at gamma-ray energies: the hallmark line at
511~keV, and the unique three-photon positronium (Ps) continuum
emission (cf.\ \cite[Guessoum, Jean, \& Gillard 2005]{Guessoum05}). 
%Direct annihilation of positrons with electrons, and their
%annihilation via the formation of para-Ps (with the spins of electron
%and positron being anti-parallel), result in the emission of two
%511~keV photons. Annihilation via the formation of ortho-Ps (with the
%spins of electron and positron being parallel) produces three photons
%and gives rise to the Ps continuum emission, which is roughly
%saw-tooth shaped with a peak at the maximum energy of 511~keV. 
The detailed shape of the 511~keV annihilation line, as well as the
positronium fraction (the fraction of positrons that annihilate
through Ps formation), depend on the physical properties of the
annihilation media; therefore detailed spectroscopy of the positron
annihilation can provide unique information on the annihilation media
and processes.
% (\cite[Guessoum, Jean, \& Gillard 2005]{Guessoum05}).

%Positrons can be produced through a variety of processes in our
%Galaxy. The main production processes are $\beta^+$ decay of
%radioactive nuclei, the decay of $\pi^+$s resulting from interactions
%of energetic nuclei, pair (electron-positron) production in
%$\gamma$-$\gamma$ interactions, and pair production by electrons in
%strong magnetic fields. Exotic processes, such as the annihilation or
%decay of light dark matter, have also been proposed (\cite[Boehm
%\etal\ 2004]{Boehm04}). In our Galaxy, astrophysical environments suitable
%for each of these processes exist. Radioactive nuclei are formed by
%explosive and hydrostatic nucleosynthesis in SNe, novae, hypernovae,
%WR stars, and AGB stars; they can also result from interactions of
%%highly energetic 
%cosmic-ray nuclei 
%%interacting 
%in the ISM, which also produce $\pi^+$s. Pair production in
%$\gamma$-$\gamma$ interactions occurs in regions of high fluxes of
%energetic photons, e.g.\ close to compact objects or in gamma-ray
%bursts. Electrons moving in very high magnetic fields can be found
%e.g.\ in the magnetospheres of pulsars. The annihilation of light dark
%matter does
%%Exotic processes usually do 
%not depend on any specific astrophysical setting, but 
%%e.g.\ the annihilation of light dark matter
%only on the light dark matter particle density. 
%
%
Positrons can be produced through a variety of processes and by a
variety of objects in our Galaxy. The main production processes are
$\beta^+$ decay of radioactive nuclei (formed in SNe, novae, hypernovae, WR
stars, and AGB stars), the decay of $\pi^+$s resulting from
interactions of energetic nuclei (in the ISM), pair (electron-positron)
production in $\gamma$-$\gamma$ interactions (close to compact objects and
in gamma-ray bursts), and pair production by electrons in strong magnetic
fields (around pulsars). Exotic processes, such as the annihilation or decay
of light dark matter, have also been proposed (\cite[Boehm
\etal\ 2004]{Boehm04}). 
%In our Galaxy, astrophysical environments suitable
%for each of these processes exist. Radioactive nuclei are formed by
%explosive and hydrostatic nucleosynthesis in SNe, novae, hypernovae,
%WR stars, and AGB stars; they can also result from interactions of
%%highly energetic 
%cosmic-ray nuclei 
%%interacting 
%in the ISM, which also produce $\pi^+$s. Pair production in
%$\gamma$-$\gamma$ interactions occurs in regions of high fluxes of
%energetic photons, e.g.\ close to compact objects or in gamma-ray
%bursts. Electrons moving in very high magnetic fields can be found
%e.g.\ in the magnetospheres of pulsars. The annihilation of light dark
%matter does
%%Exotic processes usually do 
%not depend on any specific astrophysical setting, but 
%%e.g.\ the annihilation of light dark matter
%only on the light dark matter particle density.
%
%
In spite of this
well developed understanding of the fundamental physics of positron
production, theoretical models for the positron yield of the various
sources are still highly uncertain due to uncertainties concerning the
astrophysical conditions in the different sources, and 
%concerning the sources' 
their Galactic distribution and duty cycle. 
%Obviously, more theoretical work is needed, but 
Considering also the limited quality of existing observations, it is
not surprising that the origin of the positrons is still poorly
understood.  At this stage improved measurements of the spatial
distribution and the spectrum of the Galactic annihilation radiation
are crucial for constraining better the many models.

\subsubsection{Annihilation radiation: Galatic
distribution}\label{pos_spatial} 

%\begin{figure}
%\centerline{\includegraphics[width=11cm]{juergen511_em.eps}}
%%\caption{TBD.}
%%\label{knoedlseder_511fig}
%%\end{figure}
%%
%\vspace*{2ex}
%%
%%\begin{figure}
%\centerline{\includegraphics[width=11cm,bbllx=57pt,bblly=369pt,bburx=508pt,bbury=593pt,clip=]{/users-data/weiden/SPI/PosCont/obs/All0019-0130-pub041210+vela/images/mrem2.ellipse_1.25_-0.75_10_6.CrabCygX1.ds1-f0.od.d.d.box5-5.bulge/aitoff_iter8.ps}}
%\caption{Sky maps in the 511~keV line (top panel,
%\cite{Knoedlseder05}) and the Ps continuum emission (bottom panel,
%\cite{Weidenspointner05}). Details are given in the text.}
%%\label{weidenspointner_poscntfig}
%\label{511_poscnt_fig}
%\end{figure}

\begin{figure}[t]
\begin{minipage}[b]{8.cm}
\includegraphics[width=8cm]{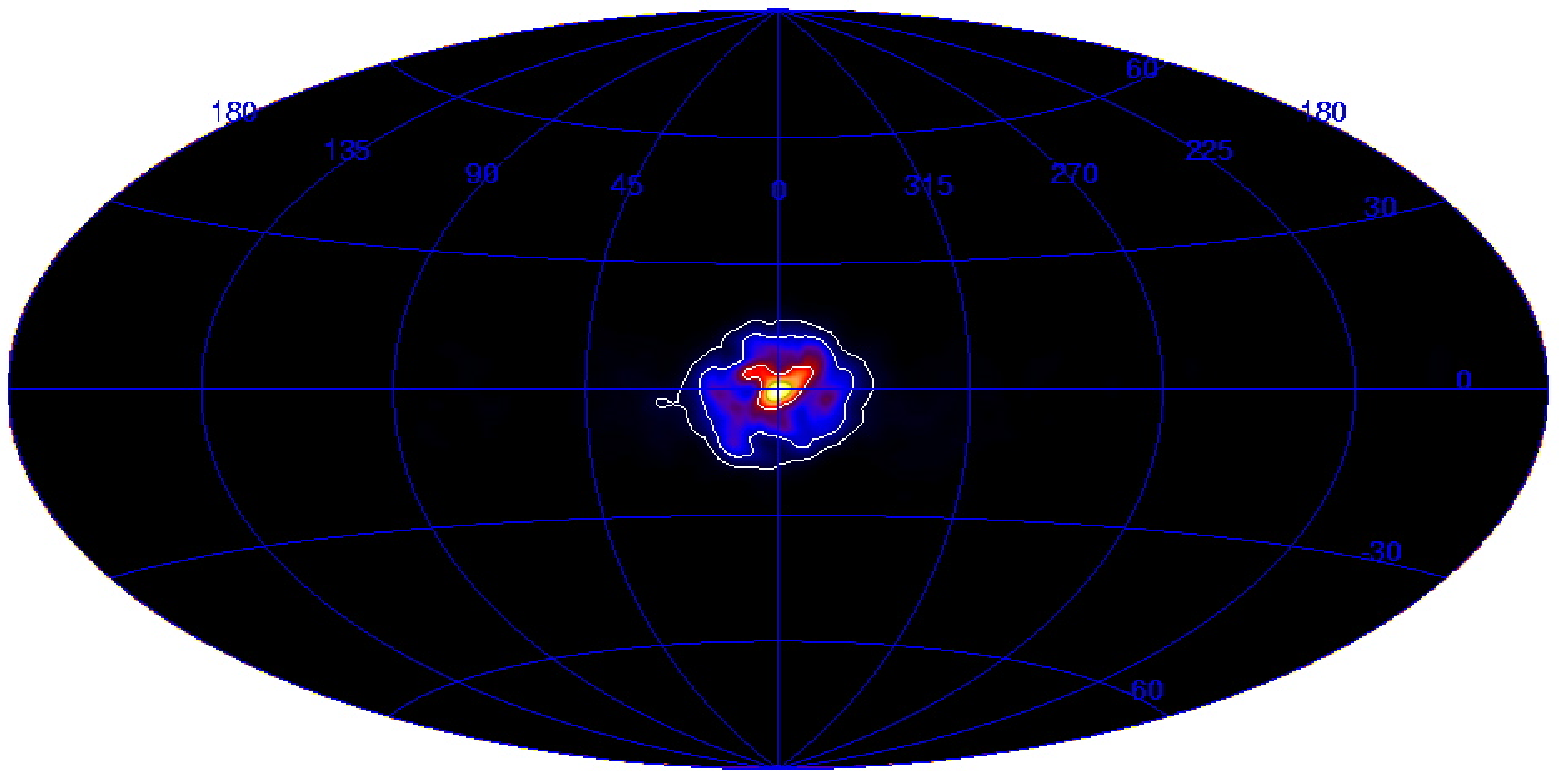}
\vspace*{2ex}
\includegraphics[width=8cm,bbllx=20pt,bblly=282pt,bburx=472pt,bbury=510pt,clip=]{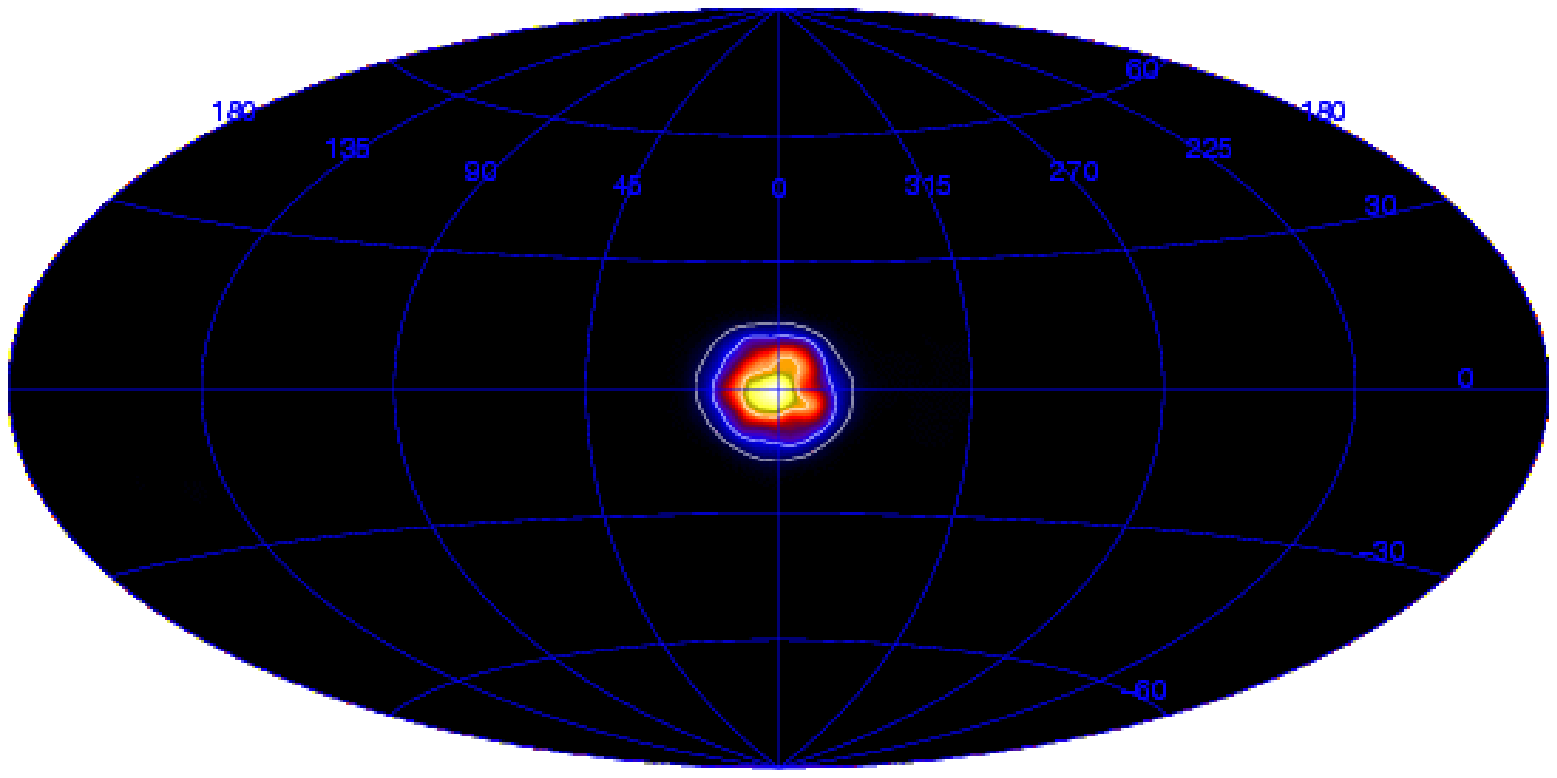}
\end{minipage}
\hfill
%\begin{minipage}[t]{4.0cm}
%\vspace*{-9.3cm}
\begin{minipage}[t]{5.0cm}
\vspace*{-8.3cm}
\caption{Sky maps in the 511~keV line (top panel,
\cite[Kn\"odlseder \etal\ 2005]{Knoedlseder05}) and the Ps continuum
emission (bottom panel, \cite[Weidenspointner \etal\
2005]{Weidenspointner05}). The maps were generated employing an
implementation of the Richardson-Lucy algorithm. To reduce noise
artifacts, the iterative corrections were smoothed during image
reconstruction with a $5^\circ \times 5^\circ$ boxcar average. Further
details are given in the text.}
\label{511_poscnt_fig}
\end{minipage}
\end{figure}

Investigations of the sky distribution of the annihilation radiation
promise to provide clues to the identification of the source(s) of
positrons in our Galaxy.
%, despite the fact that positrons may travel
%from their birth places before annihilating. 
First maps of the annihilation radiation, limited to the inner regions
of our Galaxy, were obtained using the OSSE instrument on board the
Compton Gamma-Ray Observatory in the 511~keV line and in Ps continuum
emission (e.g.\ \cite[Purcel \etal\ 1997]{Purcell97}, \cite[Milne
\etal\ 2001]{Milne01}).
%Furthermore, the
%OSSE instrument allowed \cite{Kinzer99, Kinzer01} to study the
%one-dimensional distribution in longitude and in latitude of diffuse
%emission, including annihilation radiation, from the inner Galaxy.

Improved mapping of the 511~keV line is feasible with the
commissioning of SPI.
%the imaging spectrometer SPI on board ESA's INTEGRAL observatory. 
%\citep{Jean03a, Weidenspointner04, Knoedlseder05}.  A first search for
%large-scale 511~keV line emission from the Galactic plane using SPI
%was presented by \citet{Teegarden05}.
During the first year of the mission most of the celestial sphere was
observed.
%, with the exposure peaking in the inner Galaxy. 
Using these
observations, first sky maps
%\footnote{The maps were generated
%employing an implementation of the Richardson-Lucy algorithm. To
%reduce noise artifacts, the iterative corrections were smoothed during
%image reconstruction with a $5^\circ \times 5^\circ$ boxcar average.}
in the 511~keV positron annihilation line (\cite[Kn\"odlseder \etal\
2005]{Knoedlseder05}) and in the Ps continuum emission
(\cite[Weidenspointner \etal\ 2005]{Weidenspointner05}) have been
obtained with the SPI spectrometer
%, depicted in the top and bottom panels of 
(Fig.~\ref{511_poscnt_fig}, top and bottom). In both maps of positron
annihilation radiation, the only prominent signal seen is that from
the Galactic bulge region. The surface brightness of any emission from
any other sky regions, in particular from the Galactic disk, is much
fainter. The emission appears to be symmetric about the Galactic
center (GC), and its
centroid coincides well with the GC. The differences between the two
maps, as well as any apparent small scale sub-structure in the maps,
are not significant, as will be demonstrated below.
%, they arise because the different signal-to-noise ratios in the two
%maps result  

%Mapping in principle provides an unbiased view of the sky, but has
%limitations e.g.\ for faint emission, as is the case for the Galactic
%disk. 
A more quantitative approach for studying the Galactic
distribution of the observed extended emission is model fitting. 
%
%We first modelled the positron annihilation radiation by an ellipsoidal
%distribution with a Gaussian radial profile and determined the
%best-fit centroid location and extent in Galactic longitude and
%latitude.
%
From modeling the positron annihilation radiation by an ellipsoidal
distribution with a Gaussian radial profile it can be concluded that
the annihilation radiation is spherically symmetric with a FWHM of
about $8^\circ$ and centered at the GC (\cite[Kn\"odlseder \etal\
2005]{Knoedlseder05}, \cite[Weidenspointner
\etal\ 2005]{Weidenspointner05}).
The annihilation line signal is strong
enough to allow more detailed studies of its Galactic
distribution. \cite{Knoedlseder05} fitted a
variety of Galactic bulge, halo, and disk models to the data. Stellar
bulge and halo models describe the annihilation line emission from the
central region of our Galaxy equally well. On adding a Galactic disk
component to the model fits, the faint annihilation radiation from the
Galactic disk is detected at the 3-4$\sigma$ level.

%The 511~keV line all-sky fluxes obtained for the best fitting bulge
%and halo models are $(1.05 \pm 0.06) \times
%10^{-3}$~ph~cm$^{-2}$~s$^{-1}$ and $(1.6
%\pm 0.5) \times 10^{-3}$~ph~cm$^{-2}$~s$^{-1}$, respectively. When
%combined with either of these two models, the flux attributed to a
%Galactic disk component is about $0.7 \times
%10^{-3}$~ph~cm$^{-2}$~s$^{-1}$. The total 511~keV line flux from our
%Galaxy is found to be $(1.5-2.9)
%\times 10^{-3}$~ph~cm$^{-2}$~s$^{-1}$ \footnote{Since bulge and halo
%components are alternative models their contributions should not be
%added to derive the total flux.}. Assuming a Ps fraction of 0.93
%%, in agreement with the most recent SPI results as described below, 
%this flux corresponds to a Galactic positron annihilation rate of
%$(1.6-2.8) \times 10^{43}$~s$^{-1}$. Comparing the fluxes attributed
%to the bulge or halo components to those attributed to the disk, it
%follows that the bulge-to-disk flux ratio is about 1--3 and that the
%bulge-to-disk annihilation luminosity ratio is even larger with a
%range of 3--9. Consequently, the model fits confirm the mapping
%result: SPI observations demonstrate for the first time that the
%annihilation radiation from our Galaxy is dominated by the bulge
%region (\cite{Knoedlseder05}).
%
From these model fits a total 511~keV line flux from our Galaxy of
$(1.5-2.9) \times 10^{-3}$~ph~cm$^{-2}$~s$^{-1}$ is derived, with the
bulge-to-disk flux ratio being about 1--3 and the bulge-to-disk
annihilation luminosity ratio being even larger with a range of
3--9. Consequently, the model fits confirm the qualitative mapping
result: SPI observations demonstrate for the first time that the
annihilation radiation from our Galaxy is dominated by the bulge
region (\cite[Kn\"odlseder \etal\ 2005]{Knoedlseder05}). Assuming a Ps
fraction of 0.93
%, in agreement with the most recent SPI results as described below, 
the total 511~keV line flux corresponds to a Galactic positron
annihilation rate of $(1.6-2.8) \times 10^{43}$~s$^{-1}$.

Another approach to identifying the sources of Galactic positrons is
to compare the sky distribution of the 511~keV line emission with
all-sky intensity distributions observed at other wavelengths.
%When comparing the sky distribution of the 511~keV line emission with
%known Galactic distributions, 
The relatively best agreement is found at wavelengths dominated by
emission from members of old stellar populations, but none of the
tracer maps provides an acceptable fit to the data.
% such as K and M stars or low-mass X-ray binaries. 
The sky distribution of positron annihilation appears to be
unique, as it is even more bulge dominated than potential old positron
source populations such as Type Ia supernovae, novae, or low-mass
X-ray binaries (\cite[Kn\"odlseder \etal\ 2005]{Knoedlseder05}).

%So far, it had been assumed that the emission is diffuse. 
Both
\cite{Knoedlseder05} and \cite{Weidenspointner05} searched for
evidence for contributions from point sources on top of the extended
emission, but none could be found in the 511~keV annihilation line and
the Ps continuum emission. 
%Upper flux limits down to typically
%$10^{-4}$~ph~cm$^{-2}$~s$^{-1}$ were obtained for point sources 
%%fitted in addition to diffuse emission models 
%in the 511~keV line. 
Similarly, \cite{DeCesare04} did not find any evidence for point
sources of 511~keV line emission in the GC region using observations
by IBIS.

%{\bf !!! TBD: implications: old stellar population etc. !!!}

\subsubsection{Annihilation radiation: spectroscopy}\label{pos_spec}

\begin{figure}[t]
\begin{minipage}[b]{7.cm}
\includegraphics[width=8cm]{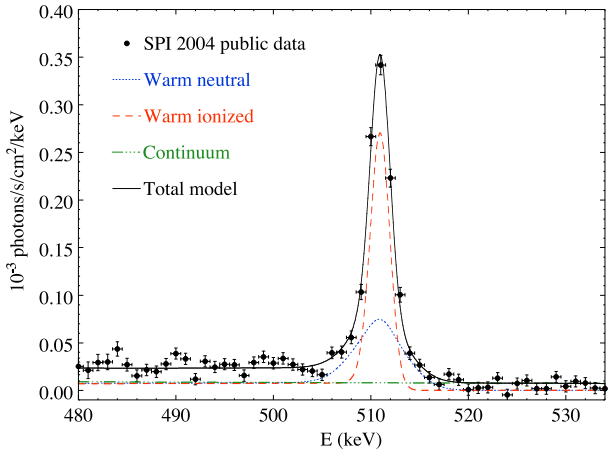}
\end{minipage}
\hfill
\begin{minipage}[t]{5.0cm}
\vspace*{-6cm}
\caption{Best fit of the SPI spectrum of the Galactic bulge with the
warm components of the ISM and the Galactic continuum emission (from
\cite[Jean \etal\ 2005]{Jean05}).} 
\label{jean_annspc_fig}
\end{minipage}
\end{figure}

The most detailed spectroscopy to date of the annihilation radiation from the
GC region has been performed by \cite{Jean05}.
% using same SPI observations that have been used for investigating the sky
%distribution. 
The line is found to be composed of a narrow and a broad component,
consistent respectively with the expected line width in a warm ISM for
Ps formation by radiative recombination and Ps formation in
flight. This result is consistent with an earlier analysis by
\cite{Churazov04}. 
When fitting the annihilation spectra that are
expected for the five standard phases of the ISM to the SPI spectrum
(see Fig.~\ref{jean_annspc_fig}), it follows that about half of the
positrons annihilate in each of the warm neutral and the warm
ionized phases. Possible fitted contributions from annihilations in
cold gas, molecular clouds, or hot gas are not significant so far. The
importance of annihilations in the warm phases is consistent with our
current understanding of positron propagation and annihilation and the
gas distribution in the Galactic bulge (\cite[Jean \etal\ 2005]{Jean05}).
\cite{Jean05} find a value for the Ps fraction of
$0.97\pm0.02$, consistent with pre-INTEGRAL results, and with the SPI
results obtained by \cite{Churazov04} and \cite{Weidenspointner05}.

%\begin{figure}
%\centerline{\includegraphics[width=9cm]{fig-fit-frac.eps}}
%\caption{Best fit of the SPI spectrum of the Galactic bulge with the
%warm components of the ISM and the Galactic continuum emission (from
%\cite{Jean05}).} 
%\label{jean_annspc_fig}
%\end{figure}

%When fitting the annihilation spectra that are expected for the five
%standard phases of the ISM to the SPI spectrum (see
%Fig.~\ref{jean_annspc_fig}), it follows that about half of the
%positrons annihilate in each the warm neutral and the warm ionized
%phase. Possible contributions from annihilations in cold gas,
%molecular clouds, or hot gas are not yet significant. The dominance of
%annihilations in the warm phases is consistent with our current
%understanding of positron propagation and annihilation and the gas
%distribution in the Galactic bulge. 

%{\bf !!! TBD: implications}

%=======================================================================

%\section{Summary and prospects}
\section{Prospects}

The prospects for further nucleosynthesis studies with INTEGRAL are
bright. 
%In many areas, the first year of observations have already
%yielded results of unprecedented quality. Among them the first sky
%maps of positron annihilation radiation, establishing the bulge region
%as dominant annihilation region, the first spatially resolved
%spectroscopy of $^{26}$Al, and the first significant measurement of
%$^{60}$Fe in our Galaxy. 
As the mission continues, we expect to learn much more about the
origin of positrons in our Galaxy from ever-improving mapping,
spatially resolved spectroscopy, and dedicated observations of
individual candidate sources. We also expect to study the Galactic
distribution of $^{26}$Al in detail by extracting distance information
from line shifts, to investigate the Galactic distribution of
$^{60}$Fe which combined with studies of $^{26}$Al will provide new
insights into massive star nucleosynthesis as well as the Galactic
star formation and SN rates, and to clarify the explosion physics of
ccSNe by observing selected SNRs like Cas~A.

%=======================================================================

%\begin{acknowledgments}
%
%{\bf !!! TBD !!!}
%
%\end{acknowledgments}

%=======================================================================

\begin{discussion}

\discuss{V\"olk}{Regarding $^{44}$Ti sources you said that only Cas~A
has been detected by INTEGRAL. Concerning the consistent claims of the
past, concerning $^{44}$Ti emission from the SNR Vela Jr., what is
then your position?}

\discuss{Weidenspointner}
{
%%First of all, I would like to emphasize that
%%past claims regarding $^{44}$Ti emission from the Vela Jr.\ SNR, at
%%least those made based on COMPTEL measurements, were not
%%conclusive. As far as INTEGRAL is concerned, the first upper limits on
%%the $^{44}$Ti line flux are based on
%%observations taken during cycle~1. Unfortunately, an intense solar
%%flare, and the failure of one the Ge detectors, resulted in only about
%%half of the data being used in the analysis. 
%%The upper limits are not
%%yet constraining previous COMPTEL results. 
%First upper limits based on observations taken during cycle~1 are not
%yet constraining previous COMPTEL results. By the end of cycle~3 the
%exposure to Vela Jr.\ will have increased by more than a factor of 3
%though, and then INTEGRAL should be able to provide a better answer to
%this intriguing question.
%
First INTEGRAL upper limits based on observations taken during Cycle~1
are still less constraining than previous COMPTEL results. By the end
of Cycle~3 the exposure to Vela Jr.\ will have increased by more than
a factor of 3 though, and then INTEGRAL should be able to provide a
better answer to this intriguing question.
}

\discuss{Bosch-Ramon}
{Is the line e-e+ of the great annihilator completely ruled out?}

\discuss{Weidenspointner}
{If the Great Annihilator is manifesting itself as a point source, or
point-like source, of narrow 511~keV line emission, then it can indeed
be completely ruled out.}

\discuss{Elvis}
{The Ps and 511~keV maps appeared to show structure.  Are these 
significant?}

\discuss{Weidenspointner}
{No. We have not yet found any significant structure on angular
scales of a few degress or below.}

\discuss{Elvis}
{The $^{60}$Fe line had similar sized features at slightly higher and
lower energies, which suggests that $^{60}$Fe may not be a significant
feature. If treated as an upper limit does this solve the Fe/Al ratio
problem?}

\discuss{Weidenspointner}
{One should keep in mind that the combined significance of the two
high-energy $^{60}$Fe lines is only 3.4$\sigma$. The feature above the
overlayed $^{60}$Fe lines is the residual of a well known background
line from $^{69}$Ge decay in the detectors, which our analysis reduces
by about a factor of 100. The weak residual below the $^{60}$Fe lines
cannot be attributed to a single instrumental background feature. If
the detected flux is treated as an upper limit, the need for a source
of $^{26}$Al in addition to ccSNe is even exacerbated.}

\end{discussion}

\end{document}